\documentclass{article}
\usepackage{fullpage,tikz,amsmath,amsfonts,amssymb}

\begin{document}
\title{\Large\bf On the Correctness of Inverted Index Based Public-Key Searchable Encryption Scheme for Multi-time Search}
\author{Shiyu Ji \\shiyu@cs.ucsb.edu}
\date{}
\maketitle

\begin{abstract}
In this short note we argue that the state-of-art inverted index based public key searchable encryption scheme proposed by Wang et al may \emph{not} be completely correct by giving a counterexample.
\end{abstract}

\section{Introduction}
In INFOCOM 2015, Wang et al \cite{Wang15} proposed the first practical inverted index based public-key searchable encryption scheme with support for multi-time search. In their paper, the correctness of the encryption scheme is argued in Theorem 1, which states their scheme returns all the documents that contain the query keyword(s). We point out if there are more than one keywords in the query, the scheme in \cite{Wang15} is possible to return a document which does not contain all the queried keywords. We show this point by giving a simple counterexample.

\newcommand{\I}{\mathcal{I}}
\section{The Counterexample}
In this section all notations and definitions are from \cite{Wang15}. Please refer to \cite{Wang15} for details.
\begin{itemize}
\item {\bf Setup}($k$): Suppose $k=4$ (the parameters we choose here are too small to be secure enough, but the correctness should also hold), and the data owner chooses two $k$-bit primes: $p=11$ and $q=13$. Then the data owner generates the key pair $(pk,sk)$ for the Paillier algorithm, in which $n = pq = 143$. The data owner also chooses the following parameters: a pseudorandom permutation $f$ mapping keyword/document identifiers to tags (in integers), and an invertible matrix $M$. For simplicity, we assume all the keyword/document identifiers are integers. Again, no matter which parameter is chosen, the correctness should be preserved.
\item {\bf IndexGen}($M$, $\I$): Suppose the inverted index $\I$ is given as follows:
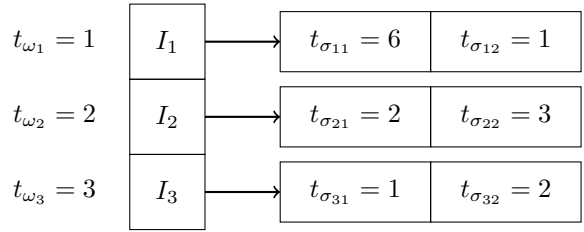
\begin{quote}
\centering
	\begin{tikzpicture}
		\node at (-1,.5) {$t_{\omega_3} = 3$}; 
		\draw (0,0) rectangle (1,1);
		\node at (.5,.5) {$I_3$};
		\draw [thick, ->] (1,.5) -- (2,.5);
		
		\node at (-1,1.5) {$t_{\omega_2} = 2$}; 
		\draw (0,1) rectangle (1,2);
		\node at (.5,1.5) {$I_2$};
		\draw [thick, ->] (1,1.5) -- (2,1.5);
		
		\node at (-1,2.5) {$t_{\omega_1} = 1$}; 
		\draw (0,2) rectangle (1,3);
		\node at (.5,2.5) {$I_1$};
		\draw [thick, ->] (1,2.5) -- (2,2.5);
		
		\draw (2,2.1) rectangle (4,2.9);
		\node at (3,2.5) {$t_{\sigma_{11}} = 6$};
		\draw (4,2.1) rectangle (6,2.9);
		\node at (5,2.5) {$t_{\sigma_{12}} = 1$};
		
		\draw (2,1.1) rectangle (4,1.9);
		\node at (3,1.5) {$t_{\sigma_{21}} = 2$};
		\draw (4,1.1) rectangle (6,1.9);
		\node at (5,1.5) {$t_{\sigma_{22}} = 3$};
		
		\draw (2,.1) rectangle (4,.9);
		\node at (3,.5) {$t_{\sigma_{31}} = 1$};
		\draw (4,.1) rectangle (6,.9);
		\node at (5,.5) {$t_{\sigma_{32}} = 2$};
	\end{tikzpicture}
\end{quote}
Here $t_{\sigma_{ij}} = f(\sigma_{ij})$, and $t_{\omega_i} = f(\omega_i) = i$ for each $i \in \{1,2,3\}$. Also we have 
$$f(\Omega) = \{1,2,3\},\quad f(\Sigma) = \{1,2,3,6\},$$
where $\Omega$ is the collection of keywords, and $\Sigma$ is the collection of documents.

The maximum length of all the inverted lists above is $L = 2$. Since each inverted list has the same length 2, the data owner does not generate any random number for each list. Then the polynomials for the lists can be obtained as follows:
$$P_1(x) = (x-6)(x-1) = x^2-7x+6.$$
$$P_2(x) = (x-2)(x-3) = x^2-5x+6.$$
$$P_3(x) = (x-1)(x-2) = x^2-3x+2.$$
Then the data owner immediately has the polynomial vectors:
$$I_1 = (1, -7, 6)^T.$$
$$I_2 = (1, -5, 6)^T.$$
$$I_3 = (1, -3, 2)^T.$$
Let $I$ be $(I_1,I_2,I_3)^T$. The data owner encrypts the coefficients in $I$ by using Paillier algorithm to get the encrypted index as $\tilde{I}$. Since $t_{\omega_i} = i$ when $i$ is 1, 2 or 3, we have the dictionary matrix:
$$M_D =\begin{pmatrix}
 1 & 8 & 27 \\
 1 & 4 & 9\\
 1 & 2 & 3
\end{pmatrix}$$
The data owner outsources $M_D' = M\cdot M_D$ and $\tilde{I}$ to the cloud.

\newcommand{\Q}{\mathcal{Q}}
\newcommand{\T}{\mathcal{T}}
\item {\bf TrapdoorGen}($M$,$\Q$): Suppose the query is $\Q = \{1,3\}$, i.e., to query the intersection of $I_1$ and $I_3$. The search user generates two random numbers $r_1$ and $r_2$ from $\mathbb{Z}_n^*$ such that neither of them is 1, 2 or 3. We will discuss the choice of $r_1$, $r_2$ later. Now the user can get the query polynomial as follows:
$$P_{\Q}'(x) = (x-2)(x-r_1)(x-r_2) = x^3-(r_1+r_2+2)x^2+(r_1r_2+2r_1+2r_2)x-2r_1r_2.$$
Thus its coefficient vector is 
$$(1, -(r_1+r_2+2), r_1r_2+2r_1+2r_2, -2r_1r_2).$$
Then the user sends the trapdoor to the cloud:
$$\T_{\Q} = ((1, -(r_1+r_2+2), r_1r_2+2r_1+2r_2)\cdot M^{-1}, Enc_{pk}(-2r_1r_2)).$$

\newcommand{\V}{\mathcal{V}}
\item {\bf Query}($\tilde{I}$,$\T_{\Q}$): The cloud server computes
$$\V = \T_{\Q}[1] \cdot M_D' = (1, -(r_1+r_2+2), r_1r_2+2r_1+2r_2)\cdot M_D.$$
Substituting $M_D$ as before, we have
$$\V = (r_1r_2+r_1+r_2-1, 2r_1r_2, 3r_1r_2-3r_1-3r_2+9).$$
Then the cloud server computes $v_i' = Enc_{pk}(v_i)+_h \T_{\Q}[2]$, where $+_h$ denotes homomorphic addition in ciphertexts (Paillier is a partially homomorphic encryption scheme):
$$\V' = (v_1',v_2',v_3') = (Enc_{pk}(-r_1r_2+r_1+r_2-1), Enc_{pk}(0), Enc_{pk}(r_1r_2-3r_1-3r_2+9)).$$
Finally the server computes $P_R(x) = \V'\cdot \tilde{I}^T$ and sends the polynomial $P_R(x)$ to the search user.\footnote{Here is one doubt: Paillier encryption is \emph{not} fully homomorphic, and hence it cannot support multiplication between two ciphertexts. Thus the encrypted polynomial $P_R(x)$ cannot be directly computed by the cloud server. One possible solution here is: the server sends $\V'$ and $\tilde{I}$ to the user and let the user decrypt them and compute the polynomial $P_R'(x)$ in plaintext.}

\item {\bf OT}($P_R$): Upon receiving $P_R(x)$, the user decrypts the coefficients in $P_R(x)$ and restores the polynomial as follows:
$$P_R'(x) = Dec_{sk}(v_1') \cdot P_1(x) + Dec_{sk}(v_3')\cdot P_3(x) = (-2r_1-2r_2+8)x^2+(4r_1r_2+2r_1+2r_2-20)x-4r_1r_2+12.$$
Clearly $x=1$ is one root of $P_R'(x)=0$. However note that it admits another root:
$$P_R'(x) = 2(x-1)((-r_1-r_2+4)x+2r_1r_2-6).$$
If $r_1+r_2\not=4$, then another root is 
$$x_2 = \frac{2r_1r_2-6}{r_1+r_2-4}.$$
Suppose $r_1 =5$ and $r_2 = 7$, then $x_2 = 8$ will also be admitted as a root (document tag) by the search user. However document tag $8$ is not in either $I_1$ or $I_3$. Thus the result is incorrect. In fact, there are many other choices of $r_1$, $r_2$ that also give incorrect roots. Here we list a few of them:
\begin{quote}
\centering
\begin{tabular}{|c | c | c|}
\hline
$r_1$ & $r_2$ & roots of $P_R'(x)$ \\
\hline
4 & 6 & 1, 7 \\
6 & 8 & 1, 9 \\
7 & 9 & 1, 10 \\
5 & 15 & 1, 9 \\
9 & 19 & 1, 14 \\
9 & 27 & 1, 15 \\
\hline
\end{tabular}
\end{quote}
Note that since $n=143$, all the $r_1$ and $r_2$ above are in $\mathbb{Z}_n^*$. All the incorrect roots are integers, and thus the search user has no way to distinguish them as redundant roots.

\end{itemize}

\section{Remarks}
The counterexample above is not surprising since given two polynomials $P_1$ and $P_2$, the roots of their sum $P_1+P_2$ include the common roots of $P_1$ and $P_2$, but may not \emph{only} contain them. In the proof of Theorem 1 in \cite{Wang15}, it is not explained why the roots of Eq. (2) are \emph{exactly} the common tags in the queried lists.

If the user only queries one keyword each time, then the scheme in \cite{Wang15} should be correct, since only one list $I_i$ is returned from the cloud server in the form of $s\cdot I_i$, where $s$ is some scalar. There is no summation of polynomials and hence the roots are exactly the tags in $I_i$.

As a possible way to repair the scheme in \cite{Wang15} for multiple keyword search, the user may query the same keywords for several times, and take the intersection of the returned document tags. However this is more time-consuming, and it is not clear how many query times are enough to guarantee the correctness.


\begin{thebibliography}{19}
\bibitem{Wang15}
Wang, B., Song, W., Lou, W. and Hou, Y.T., ``Inverted index based multi-keyword public-key searchable encryption with strong privacy guarantee''. In 2015 IEEE Conference on Computer Communications (INFOCOM) (pp. 2092-2100).

\end{thebibliography}
\end{document}